 \documentclass[manuscript]{emulateapj}

 \usepackage{apjfonts}

\slugcomment{to appear in the Astrophysical Journal}

\shorttitle{Progenitors of Type Ia Supernovae}
\shortauthors{Hachisu et al.}

\begin{document}

\title{Young and Massive Binary Progenitors of Type I\lowercase{a}
Supernovae \\ and Their Circumstellar Matter}


\author{Izumi Hachisu}
\affil{Department of Earth Science and Astronomy,
College of Arts and Sciences, University of Tokyo,
Komaba 3-8-1, Meguro-ku, Tokyo 153-8902, Japan}
\email{hachisu@ea.c.u-tokyo.ac.jp}

\author{Mariko Kato}
\affil{Department of Astronomy, Keio University,
Hiyoshi 4-1-1, Kouhoku-ku, Yokohama 223-8521, Japan}
\email{mariko@educ.cc.keio.ac.jp}

\and

\author{Ken'ichi Nomoto}
\affil{Department of Astronomy, University of Tokyo, Hongo 7-3-1,
Bunkyo-ku, Tokyo 113-0033, and
Institute for the Physics and Mathematics of the Universe,
University of Tokyo, Kashiwa, Chiba 277-8582, Japan}
\email{nomoto@astron.s.u-tokyo.ac.jp}



\begin{abstract}
We present new evolutionary models for Type Ia supernova (SN Ia)
progenitors, introducing mass-stripping effect on
a main-sequence (MS) or slightly evolved companion star by winds
from a mass-accreting white dwarf (WD).
The mass-stripping attenuates the rate of mass transfer from the 
companion to the WD.  As a result, quite a massive MS companion
can avoid forming a common envelope and increase the WD mass up to
the SN Ia explosion.  Including the mass-stripping effect,
we follow binary evolutions of various WD + MS systems
and obtain the parameter region in the initial donor mass -- orbital
period plane where SNe Ia occur.
The newly obtained SN Ia region extends to donor masses of
$6-7 ~M_\sun$, although its extension depends on the efficiency
of mass-stripping effect.  The stripped matter would mainly be
distributed on the orbital plane and form very massive circumstellar
matter (CSM) around the SN Ia progenitor.
It can explain massive CSM around
SNe Ia/IIn(IIa) 2002ic and 2005gj as well as tenuous CSM
around normal SN Ia 2006X.
Our new model suggests the presence of very young ($\lesssim 10^8$~yr)
populations of SNe Ia, being consistent
with recent observational indications of young population SNe Ia. 
\end{abstract}




\section{Introduction}
The nature of Type Ia supernova (SN Ia) progenitors has not been
clarified yet \citep[e.g.,][]{nie04, nom00}, although it has been
commonly agreed that the exploding star is a mass-accreting carbon-oxygen
white dwarf (C+O WD).  For the exploding WD itself, the observed
features of SNe Ia are better explained by the Chandrasekhar mass
model than the sub-Chandrasekhar mass model \citep[e.g.,][]{liv00}.
However, there has been no clear observational indication
as to how the WD mass gets close enough to the Chandrasekhar mass for
carbon ignition; i.e., whether the WD accretes H/He-rich matter from
its binary companion [single degenerate (SD) scenario], or two C+O WDs
merge [double degenerate (DD) scenario].

Recently, the following two important findings have been reported in
relation to the SN Ia progenitors: (1) circumstellar matter (CSM)
around the progenitors, and (2) a very young ($\lesssim 10^8$~yr)
population of the progenitors.

{\bf Circumstellar Matter:} In the SD scenario, H/He-rich CSM is
expected to exist around SNe Ia as a result of mass transfer from
the companion as well as the WD winds \citep*[e.g.,][]{nom82, hkn99}.
Thus searching for H/He-rich CSM is one of the key observations
to identify the progenitors \citep[e.g.,][]{lun03}.
Recently detections of such CSM have been reported for several
SNe Ia, i.e., observations of narrow H-emission lines
in SNe 2002ic \citep{hau03} and 2005gj \citep{ald06,pri07}
(Type Ia/IIn or IIa \citep{den04}), thermal X-rays from 2005ke 
\citep{imm06}, and \ion{Na}{1}~D lines in 2006X \citep{pat07a}.

The identification of SN 2002ic as an SN Ia has been confirmed by the
recent spectral comparison between SN 2005gj and SNe Ia \citep{pri07},
being against the Type Ic suggestion by \citet{ben06}.  Several CSM
interaction models suggested a $1 - 2 ~M_\sun$ CSM \citep{chu04,nom05}.  
The evolutionary origin of such a massive CSM has been explored by
\citet{liv03} based on a common envelope evolution model,
by \citet{han06} from the delayed dynamical instability model of
binary mass transfer, and by \citet{woo06} based on a recurrent
nova model with a red giant companion.

For normal SNe Ia, non-detection of radio has put the upper limit of
mass loss rate as $\dot M / v_{10} \lesssim 10^{-8} M_\sun$~yr$^{-1}$,
where $v_{10} \equiv v / 10$~km~s$^{-1}$ \citep{pan06}.
However, the optical observations of SN 2006X have
detected variable Na I D lines from CSM, whose expansion velocity and
mass have been estimated to be $v_{10} \sim$ 10 and $\sim 10^{-4}
~M_\sun$ \citep{pat07a}.  Patat et al. have suggested that the CSM in
SN 2006X originated from the red-giant companion because of relatively
low velocities.  Comparing the SN 2006X light curves with the other
normal SNe Ia light curves, \citet{wan07a} suggested that the obvious
deviation, the decline rate is slowing down in a later phase,
can be explained by an interaction between ejecta and CSM or
a light echo of circumstellar/interstellar matter 
\citep[see also][]{wan07b}.

{\bf Young Population:} According to \citet{man06}, the present
observational data of SNe Ia
are best matched by a bimodal population of the progenitors, in which
about 50 percent of SNe Ia explode soon after their stellar birth
at the {\sl delay time} of $t_{\rm delay} \sim 10^8$ yr,
while the remaining 50 percent have a much wider distribution
of the {\sl delay time} of $t_{\rm delay} \sim$ 3 Gyr.
\citet{aub07} recently reported evidence for a short
(less than 70 Myr) delay time component in the SN Ia population.
In this paper, we define the term {\sl delay time} as {\sl the
age of a binary system at the SN Ia explosion}, in order to compare
our results with the earlier results \citep[e.g.,][]{gre83, gre05, man06}.


\begin{figure*}
\epsscale{0.6}
\plotone{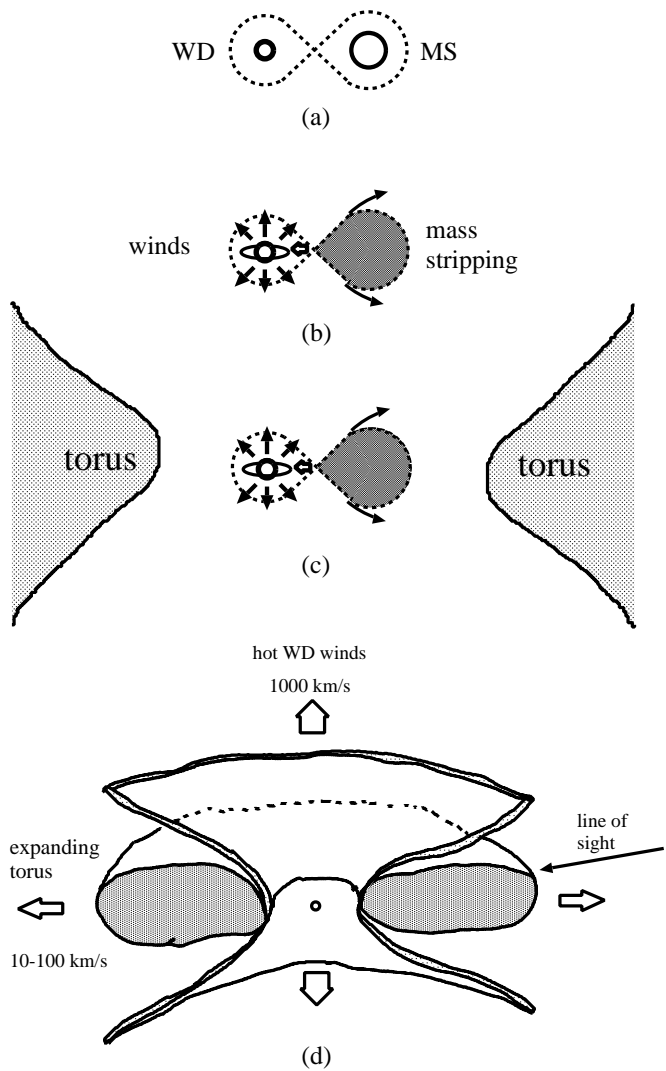}
\caption{
A schematic configuration of a binary evolution including
mass-stripping effect.
(a) Here we start a pair of a C+O WD and a more massive main-sequence
(MS) star with a separation of several to a few tens of solar radii.
(b) When the secondary evolves to fill its Roche lobe, mass transfer
onto the WD begins.  The mass transfer rate exceeds a critical rate
for optically thick winds.  Strong winds blow from the WD.
(c) The hot wind from the WD hits the secondary and strips off its surface.
(d) Such stripped-off material forms a massive circumstellar disk or
torus and it gradually expands with an outward velocity of 
$\sim 10-100$~km~s$^{-1}$.  The interaction between the WD wind
and the circumstellar torus forms an hourglass structure.  The WD mass
increases up to $M_{\rm Ia}= 1.38 ~M_\sun$ and explodes as an SN Ia. 
When we observe the SN Ia from a high inclination angle such as denoted
by ``line of sight,'' circumstellar matter (CSM) can be detected
as absorption lines like in SN 2006X. 
\label{stripping_evolution}}
\end{figure*}

This kind of short delay times ($t_{\rm delay} \lesssim 10^8$~yr) of
SNe Ia have been suggested from the distribution of
SNe Ia relative to spiral arms \citep[e.g.,][]{bar94, del94}.
Recently, \citet{dis03} reported, based on the {\it Chandra} data from
four external galaxies: an elliptical galaxy (NGC 4967),
two face-on spiral galaxies (M101 and M83), and an interacting galaxy (M51),
that in every galaxy there are at least several hundred luminous
supersoft X-ray sources (SSXSs) with a luminosity of 
$\gtrsim 10^{37}$ erg~s$^{-1}$ and that, in the spiral galaxies M101,
M83, and M51, SSXSs appear to be associated with the spiral arms.
The latter may indicate that SSXSs are young systems,
possibly younger than $10^8$ yr, and has some close relation to the
young population of SNe Ia.

The SD scenario has ever not predicted such young populations
of $t_{\rm delay} \sim 10^8$ yr, corresponding to, at least,
the zero-age main-sequence (ZAMS) stars at mass $5-6 ~M_\sun$
\citep[see, e.g.,][]{lih97, hknu99, lan00, han04}.
In the present paper, we propose a
scenario for such a young SN Ia population by introducing
{\sl mass-stripping effect} into binary evolutions. 
Mass-accreting WDs blow optically thick winds when the mass
transfer rate to the WD exceeds the critical rate of ${\dot M}_{\rm cr}
\sim 1 \times 10^{-6} M_\sun$~yr$^{-1}$ \citep{hkn96}.
The WD wind collides with the secondary's surface and strips off matter.
When the mass-stripping effect is efficient enough, the mass transfer
rate to the WD is attenuated and the binary can avoid the formation
of a common envelope even for a rather massive secondary.

The mass-stripping effect on a MS companion has been first introduced
by \citet{hac03kb, hac03kc}, who analyzed two quasi-periodic
transient supersoft X-ray sources, RX~J0513.9$-6951$ and V~Sge:
RX~J0513 shows a quasi-periodic oscillation between optical high
($\sim 100-120$ days) and low ($\sim 40$ days) states with an amplitude
of 1 mag \citep{alc96}.  RX~J0513 is X-ray bright only during
the optical low states \citep{rei00}.
\citet{hac03kb} proposed a model that the mass
transfer is modulated by the WD wind because the WD wind collides
with the companion and strips off its surface and attenuates the
mass transfer rate.  When the mass transfer rate decreases below
the critical rate ${\dot M}_{\rm cr}$, the WD wind stops and 
supersoft X-ray turns on.  This corresponds to an optical low sate.
Then the mass-transfer rate recovers because of no attenuation by
WD winds and the WD blows winds again.
X-ray turns off and an optical high state resumes and the binary
starts the next cycle of quasi-periodic oscillation.
Such a self-sustained model naturally explains major characteristics
of quasi-periodic high and low states and this success encourages us
to adopt the same idea in the evolution scenario of supersoft X-ray
sources and SN Ia progenitors.

In the present paper, we show that this mass-stripping effect derives
(1) formations of circumstellar matter (CSM) around SNe Ia and (2) a
very young population of SNe Ia.  We summarize our basic
treatments of mass-stripping effect and binary evolutions in \S 2, and
then show our numerical results and their relations to a very young
population of SNe Ia in \S 3.  In \S 4 we present the origin of
CSM around SNe Ia based on our results and show a relation between
the very young population of SNe Ia and their massive CSM.
Discussion and concluding remarks follow in \S\S 5 and 6.


\begin{figure*}
\epsscale{0.7}
\plotone{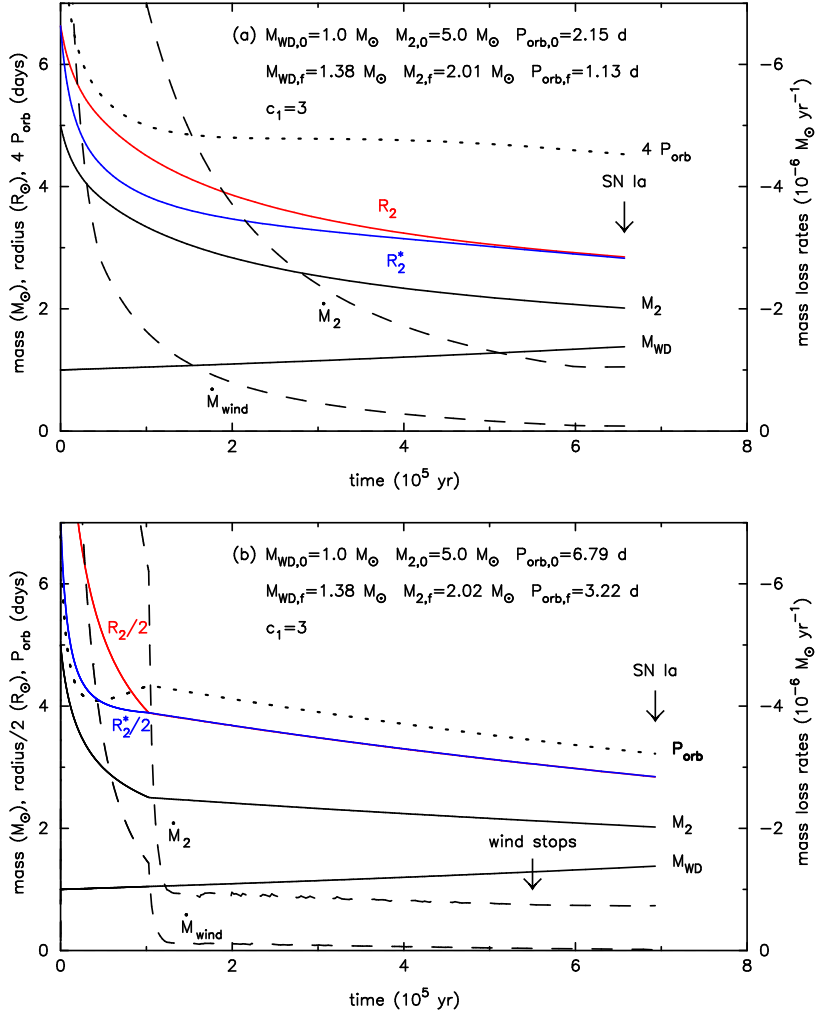}
\caption{
SN Ia evolutions for two typical cases of WIND and CALM.
(a) Case WIND: starting from $M_{\rm WD,0} =1.0 ~M_\sun$,
$M_{2,0} =5.0 ~M_\sun$, and $P_0 = 2.15$ days with $c_1=3$,
the WD reaches the SN Ia explosion in the wind phase at 
$t=6.57 \times 10^5$ yr.  The WD mass ($M_{\rm WD}$),
secondary mass ($M_2$), mass loss rate from the secondary (${\dot M}_2$),
WD wind mass loss rate (${\dot M}_{\rm wind}$), radius of the secondary
($R_2$), effective radius of the Roche lobe for the secondary
($R_2^*$), and orbital period ($P_{\rm orb}$) are plotted.
Only the orbital period is multiplied by four to easily see its change.
(b) Case CALM: starting from $M_{\rm WD,0} =1.0 ~M_\sun$,
$M_{2,0} =5.0 ~M_\sun$, and $P_0 = 6.79$ days with $c_1=3$,
the WD reaches the SN Ia explosion but in an SSXS phase without
winds at $t=  6.93 \times 10^5$ yr.
The WD wind stops at $t=5.5 \times 10^5$~yr.
Even after that, the WD loses its mass due to weak helium shell flashes
\citep{kat99h}.  Here ${\dot M}_{\rm wind}$ includes an average
mass loss rate by helium shell flashes and thus does not become
zero after the optically thick wind of steady hydrogen shell burning stops. 
Values of the secondary radius ($R_2$) and the Roche lobe radius
for the secondary ($R_2^*$) are divided by two to squeeze them
into the figure. 
\label{evolution_sn2005gj}}
\end{figure*}

\section{Mass-Stripping Effect and Binary Evolution}
Strong winds from a mass-accreting WD collide with the
companion star and strip off its surface.  This mass-stripping effect
plays an important role in binary evolutions \citep[e.g.][]{hkn99}.
Here we reformulate its treatment in our binary evolution calculation.

\subsection{New Aspects of Binary Evolutions}
First we briefly introduce a new binary evolutionary process through
stages (a)-(d) below (also shown in Fig.
\ref{stripping_evolution}a--d),
where (c) and (d) are new stages introduced by mass-stripping.

(a) The more massive (primary) component of a binary evolves to
a red giant star (with a helium core) or an AGB star (with a 
C+O core) and fills its Roche lobe.
Mass transfer from the primary to the secondary begins and a common
envelope is formed.  After the first common envelope evolution,
the separation shrinks and the primary component
becomes a helium star or a C+O WD.  The helium star
evolves to a C+O WD after a large part of helium is exhausted
by core-helium-burning.
We eventually have a close pair of a C+O WD and a main-sequence
(MS) star as shown in Figure \ref{stripping_evolution}a.

(b) After the secondary evolves to fill its Roche lobe,
the mass transfer to the WD begins.  This mass transfer occurs
in a thermal timescale because the secondary mass is more massive
than the WD.  The mass transfer rate exceeds the critical rate
for the optically thick wind to blow from the WD
\citep{hkn96, hkn99, hknu99}.

(c) Optically thick winds from the WD collide with the secondary surface
and strips off its surface layer \citep{hac03ka, hac03kb, hac03kc}.
This mass-stripping attenuates the rate of mass transfer from
the secondary to the WD, thus preventing the formation of
a common envelope for a more massive secondary in the case with 
than in the case without this effect.
Thus the mass-stripping effect widens the donor mass range of
SN Ia progenitors (see Fig. \ref{zams_strip_reg100} below).

(d) Such stripped-off matter forms a massive circumstellar
torus on the orbital plane, which may be gradually expanding with
an outward velocity of $\sim 10-100$~km~s$^{-1}$ (Fig.
\ref{stripping_evolution}d), because the escape velocity
from the secondary surface to L3 point is
$v_{\rm esc} \sim [(\phi_{\rm L3} - \phi_{\rm MS}) G M / a]^{1/2} \sim
100$~km~s$^{-1}$ (see below).  Subsequent interaction between the fast
wind from the WD and the very slowly expanding circumbinary torus forms
an hourglass structure (Fig. \ref{stripping_evolution}c--d).

\subsection{Formulation of Mass-stripping}
Fast strong winds collide with
the companion as illustrated in Figure \ref{stripping_evolution}.
The companion's surface gas is shock-heated and ablated in the wind.
We estimate the shock-heating by assuming that the velocity component
normal to the companion's surface is dissipated by the shock and
the kinetic energy is converted into the thermal energy of
the surface layer.  The heated surface layer expands to be ablated
in the wind.

  To obtain the mass stripping rate,
we use the same formulation proposed by \citet{hac03kb, hac03kc}.
We equate the stripping rate times the gravitational potential at the
companion surface to the net rate of energy dissipation by the shock as:
\begin{equation}
{{G M} \over {a}} \left( \phi_{\rm L3} - \phi_{\rm MS} \right) \cdot
{\dot M}_{\rm strip}
=  {1 \over 2} v^2 \cdot \eta_{\rm eff} \cdot g(q) \cdot {\dot M}_{\rm wind},
\label{strip_off_origin}
\end{equation}
where $M=M_{\rm WD}+ M_{\rm MS}$, $M_{\rm WD}$ is the WD mass,
$M_{\rm MS}$ is the main-sequence companion mass,
$a$ is the separation of the binary;
$\phi_{\rm MS}$ and $\phi_{\rm L3}$ denote
the Roche potential (normalized by $GM/a$)
at the MS surface and the L3 point near the MS companion, respectively;
$v$ is the WD wind velocity,
$\eta_{\rm eff}$ is the efficiency of conversion from kinetic energy
to thermal energy by the shock, $g(q)$ is the geometrical factor of
the MS surface hit by the wind including the inclination 
(oblique shock) effect of
the wind velocity against the companion's surface
\citep*[see][for more details on $g(q)$]{hkn99}, and $q \equiv M_2
/ M_1 = M_{\rm MS} / M_{\rm WD}$ is the mass ratio.  Here we modified
equation (21) of \citet{hkn99} to include the effect of Roche lobe
overflow from the L3 point.  Then the stripping rate is estimated as
\begin{equation}
{\dot M}_{\rm strip} = c_1 {\dot M}_{\rm wind},
\label{mass_stripping_rate}
\end{equation}
where
\begin{equation}
c_1 \equiv {{\eta_{\rm eff} \cdot g(q)} \over
{\phi_{\rm L3} - \phi_{\rm MS}}}
\left({{v^2 a} \over {2 G M}} \right).
\end{equation}
Here we assume that the WD wind is spherically symmetric.
If the asphericity of the WD wind is not so large, having
a latitudinal ($\theta$-angle) dependency like a broad angle jet,
we have a different form of $g(q)$
and its value may be much smaller than that for the spherically
symmetric WD winds.  We also assume $\eta_{\rm eff}=1$
in the present calculation.
When the wind velocity is as fast as 4,000 km~s$^{-1}$,
we have $c_1 \sim 10$ as estimated by \citet{hac03kb}.
Although there is a large ambiguity in this kind of parameterization
as $c_1$, \citet{hac03kb, hac03kc} found the best fit models with
$c_1= 1.5-10$ for RX~J0513.9$-6953$ and $c_1= 7-8$ for V~Sge.
We thus assume $c_1= 1$, 3, and 10 to examine the dependence
on the mass-stripping effect, because the essential ambiguity of
our formulation is included in the $c_1$ parameter.

     When winds blow from the WD and strip off the companion's
surface, the change of the separation, $\dot a$, is calculated from
\begin{eqnarray}
{{\dot a} \over a} &=& {{{\dot M}_1 + {\dot M}_2} \over {M_1+M_2}}
- 2 {{{\dot M}_1} \over {M_1}} - 2 {{{\dot M}_2} \over {M_2}}
+ 2 {{\dot J} \over {J}} \cr
&=& {{{\dot M}_1 + {\dot M}_2} \over {M_1+M_2}}
- 2 {{{\dot M}_1} \over {M_1}} - 2 {{{\dot M}_2} \over {M_2}} \cr
& & + 2 {{M_1 + M_2} \over {M_1 M_2}}
\left( \ell_{\rm w} {\dot M}_{\rm wind} + \ell_{\rm s} {\dot M}_{\rm strip}
\right),
\label{separation_decrease_by_wind}
\end{eqnarray}
where $M_1 = M_{\rm WD}$, $M_2 = M_{\rm MS}$,
$\ell_{\rm w}$ and $\ell_{\rm s}$ are
the specific angular momenta of the WD wind and the stripped-off matter,
respectively, in units of $a^2 \Omega_{\rm orb}$ with $\Omega_{\rm orb}$
being the orbital angular velocity.
Since the WD wind is much faster than the orbital motion,
the wind cannot get angular momentum from the orbital torque
during its journey, so that the wind has the same specific angular momentum
as the WD, which is estimated as
\begin{equation}
\ell_{\rm w} = \left( {{q} \over {1+q}} \right)^2.
\label{fast_wind_angular_momentum}
\end{equation}
The ablated gas from the companion is assumed to have the angular
momentum at the companion's surface.  Then
we have a numerical factor of
\begin{equation}
\ell_{\rm s} = {{h(q)} \over {g(q)}},
\label{stripping_matter_angular_momentum}
\end{equation}
which was given in Table 1 of \citet{hkn99} and is rather small
compared with $\ell_{\rm w}$.  \citep[See][for more details of
$\ell_{\rm s}$.]{hkn99}

\subsection{Modified Mass Transfer Rate}
     We have followed binary evolutions from the initial
state of $(M_{\rm 1,0}$, $M_{\rm 2,0}$, $P_0)$, i.e.,
$(M_{\rm WD,0}$, $M_{\rm MS,0}$, $P_0)$, 
where $P_0$ is the initial orbital period.
Here, the subscript naught (0) denotes stage (a) in
Figure \ref{stripping_evolution}, that is, before the mass transfer
from the secondary starts.  The radius, $R_2(M_2,t)$, and
luminosity, $L_2(M_2,t)$,
of stars which have slightly evolved off from the zero-age main-sequence
(ZAMS), are calculated using the analytic form given by \citet{tou97}.

The mass transfer proceeds on a thermal time scale
when the mass ratio $M_2/M_1$ exceeds 0.79.  We approximate
the mass transfer rate as
\begin{equation}
- {\dot M}_2  = {{M_2} \over {\tau_{\rm KH}}} \cdot \max\left(
{{\zeta_{\rm RL} - \zeta_{\rm MS}} \over {\zeta_{\rm MS}}}, 1 \right),
\label{secondary_mass_transfer}
\end{equation}
where $\tau_{\rm KH}$ is the Kelvin-Helmholtz timescale given by
\begin{equation}
\tau_{\rm KH} \approx 3 \times 10^7{\rm ~yr~}\left({{M_2} \over {M_\sun}}
\right)^2 \left({{R_2} \over {R_\sun}} \cdot 
{{L_2} \over {L_\sun}}\right)^{-1}
\label{KH_timescale}
\end{equation}
\citep[e.g.,][]{pac71b}, and $\zeta_{\rm RL}=d \log R^*/d \log M$ and
$\zeta_{\rm MS}= d \log R_{\rm MS}/d \log M$ are
the mass-radius exponents of the inner critical Roche lobe and
the main sequence component, respectively \citep[e.g.,][]{hje87}.
The effective radius of the inner critical Roche lobe, $R^*$, is
calculated from Eggleton's (1983) empirical formula, i.e.,
\begin{equation}
{{R^*} \over {a}} = f(q) \equiv  {{0.49q^{2/3}}
\over{0.6q^{2/3}   +  \ln  (1+q^{1/3})}} ,
\label{inner_critical_Roche_lobe}
\end{equation}
where $q=M_2/M_1$.

When the mass transfer rate to the WD exceeds a critical value, which
is given by
\begin{equation}
{\dot M}_{\rm cr} \approx 0.75 \times 10^{-6} \left( {{M_{\rm WD}}
\over {M_\sun}} - 0.4 \right) ~M_\sun~{\rm yr}^{-1},
\label{critical_rate_wind}
\end{equation}
for the solar composition (hydrogen content of $X=0.7$ and metallicity
of $Z=0.02$), the WD blows a wind with a mass loss rate of 
${\dot M}_{\rm wind}~(<0)$.  This critical rate of ${\dot M}_{\rm cr}$
is the same as the critical rate for mass-accreting
WDs to expand to a giant size, i.e., ${\dot M}_{\rm RG}$
\citep[see][for the recent calculation of ${\dot M}_{\rm RG}$]{nom07}.
The mass loss from the WD also occurs during the hydrogen shell flashes
when $- \dot M_2 < \dot M_{\rm stable}$, where $\dot M_{\rm stable}$
is the lowest rate for steady hydrogen burning and given by equation 
\begin{equation}
{\dot M}_{\rm stable} \approx 0.31 \times 10^{-6} \left( {{M_{\rm WD}}
\over {M_\sun}} - 0.54 \right) ~M_\sun~{\rm yr}^{-1}
\label{stable_hydrogen_burning}
\end{equation}
\citep{nom07}.  When $\dot M_{\rm stable} < - \dot M_2 <
\dot M_{\rm cr}$, we have no mass loss associated with steady hydrogen
shell-burning but have mass loss by helium shell flashes.
This mass loss play some role in the binary evolution \citep{kat99h}.
Therefore, $\dot M_{\rm wind}$ is the summation of the optically
thick wind mass loss, hydrogen shell flashes, and helium shell flashes.

We have the relation
\begin{equation}
{\dot M}_1 + {\dot M}_2 = {\dot M}_{\rm wind} + {\dot M}_{\rm strip},
\label{total_mass_conservation}
\end{equation}
from the total mass conservation, thus defining the net mass transfer
rate to the WD as
\begin{equation}
{\dot M}_{\rm transfer} \equiv {\dot M}_{\rm strip} - {\dot M}_2
= {\dot M}_1 - {\dot M}_{\rm wind},
\label{net_mass_transfer_conserve}
\end{equation}
where signs of ${\dot M}_{\rm transfer} > 0$,
${\dot M}_{\rm strip} \le 0$, ${\dot M}_2 < 0$,
${\dot M}_1 \ge 0$, and ${\dot M}_{\rm wind} \le 0$
should be noted.  If $\dot M_2$ is given, we have the net mass transfer
rate of
\begin{eqnarray}
{\dot M}_{\rm transfer} & = &
 \left\{
\begin{array}{cc}
({c_1 \dot M_{\rm cr} - {\dot M}_2})/({c_1 + 1}) ,
 & {\rm ~for~} - \dot M_2 > \dot M_{\rm cr} \\
- \dot M_2  , & {\rm ~for~} - \dot M_2 \le \dot M_{\rm cr}
\end{array}     \right. ,
\label{net_mass_transfer_on_c1}
\end{eqnarray}
where we use equations (\ref{mass_stripping_rate}),
(\ref{net_mass_transfer_conserve}), and a relation of
\begin{equation}
- {\dot M}_{\rm wind}  = 
\dot M_{\rm transfer} - \dot M_{\rm cr} ,
\end{equation}
for $- \dot M_2 > \dot M_{\rm cr}$.  Other treatments for binary
evolution are essentially the same as those in \citet{hknu99}.

Figure \ref{evolution_sn2005gj} shows two typical evolutionary sequences
that demonstrate the effects by the modified mass transfer rate,
${\dot M}_2$, in equation (\ref{secondary_mass_transfer}).

(a) Starting from $M_{\rm WD,0} =1.0 ~M_\sun$,
$M_{2,0} =5.0 ~M_\sun$, and $P_0 = 2.15$ days with $c_1=3$,
the WD reaches the SN Ia explosion in the wind phase (Case WIND)
at $t=6.57 \times 10^5$ yr after the secondary fills its Roche lobe.
The WD increases its mass
($M_{\rm WD}$) up to $M_{\rm Ia}= 1.38 ~M_\sun$ to explode as an SN Ia.
The secondary mass ($M_2$) decreases to $2.01~M_\sun$ at the explosion.
Both the mass decreasing rate of the secondary
(dashed line labeled ${\dot M}_2$) and the WD wind mass loss
rate (dashed line labeled ${\dot M}_{\rm wind}$) are also decreasing
rapidly especially in the early phase of ~$t \lesssim 1 \times 10^5$~yr. 
This is because $-{\dot M}_2$ is large and the mass transfer rate, 
${\dot M}_{\rm transfer}$, is large during this phase,
and as a result, both the WD wind mass loss rate, $\dot M_{\rm wind}$,
and the stripping rate, $\dot M_{\rm strip}$, are also large.
Shortly after this early phase,
the Roche lobe's mass-radius exponent, $\zeta_{\rm RL}$, becomes
smaller than the secondary's mass-radius exponent, $\zeta_{\rm MS}$,
that is, $\zeta_{\rm RL} - \zeta_{\rm MS} < 0$.  This gives
$- {\dot M}_2 = M_2 / \tau_{\rm KH}$ from
equation (\ref{secondary_mass_transfer}).
We keep this mass transfer rate as long as the secondary
overfills the Roche lobe, i.e., $R_2 > R_2^*$.
In Figure \ref{evolution_sn2005gj}a, we plot the secondary radius
(the red line labeled $R_2$) and the Roche lobe radius for the
secondary component (the blue line labeled $R_2^*$) to show
the condition of $R_2 > R_2^*$ during the evolution.

(b) Starting from $M_{\rm WD,0} =1.0 ~M_\sun$,
$M_{2,0} =5.0 ~M_\sun$, and $P_0 = 6.79$ days with $c_1=3$,
the WD reaches the SN Ia explosion but in a phase of no winds
(Case CALM) at $t=  6.93 \times 10^5$ yr after the secondary fills
its Roche lobe.
In this case the evolution of the mass transfer rate is different
from Case WIND above.  With $- {\dot M}_2 = M_2 / \tau_{\rm KH}$ 
for $\zeta_{\rm RL} < \zeta_{\rm MS}$ in
equation (\ref{secondary_mass_transfer}), the secondary eventually
underfills the Roche lobe, i.e., $R_2 < R_2^*$.
This can be seen in Figure \ref{evolution_sn2005gj}b, where the line
of $R_2$ crosses the line of $R_2^*$ at
$t \sim 1 \times 10^5$~yr.  This is because the stripped matter
has rather low specific angular momentum (eq.
[\ref{stripping_matter_angular_momentum}]), so that
the binary separation hardly shrinks or even 
increases as seen from the temporal increase
in the orbital period in Figure \ref{evolution_sn2005gj}b. 
In realistic binary evolutions, the mass transfer is tuned in a way
that the secondary radius is always equal to the Roche lobe radius for the
secondary, i.e., $R_2 = R_2^*$.  Therefore $- {\dot M}_2$ is 
drastically decreased after $t \sim 1 \times 10^5$~yr,
as shown in Figure \ref{evolution_sn2005gj}b.
Thus, the optically thick WD wind stops at $t=5.5 \times 10^5$~yr.
In such a low mass transfer phase as
${\dot M}_{\rm transfer} \sim 1 \times 10^{-6} M_\sun$~yr$^{-1}$, weak
helium shell flashes occur and play an important role as a mass loss
mechanism.  This helium flash wind also strips off the secondary surface,
thus working as a stripping effect.  We introduce mass-stripping effect
by these helium shell flashes into our binary evolution.
Very small but finite
${\dot M}_{\rm wind}$ in Figure \ref{evolution_sn2005gj}b (after
winds stop) represents the mass loss from the WD at helium shell flashes
and ${\dot M}_2$ includes the ensuing mass-stripping from
the secondary.


\begin{figure*}
\epsscale{0.8}
\plotone{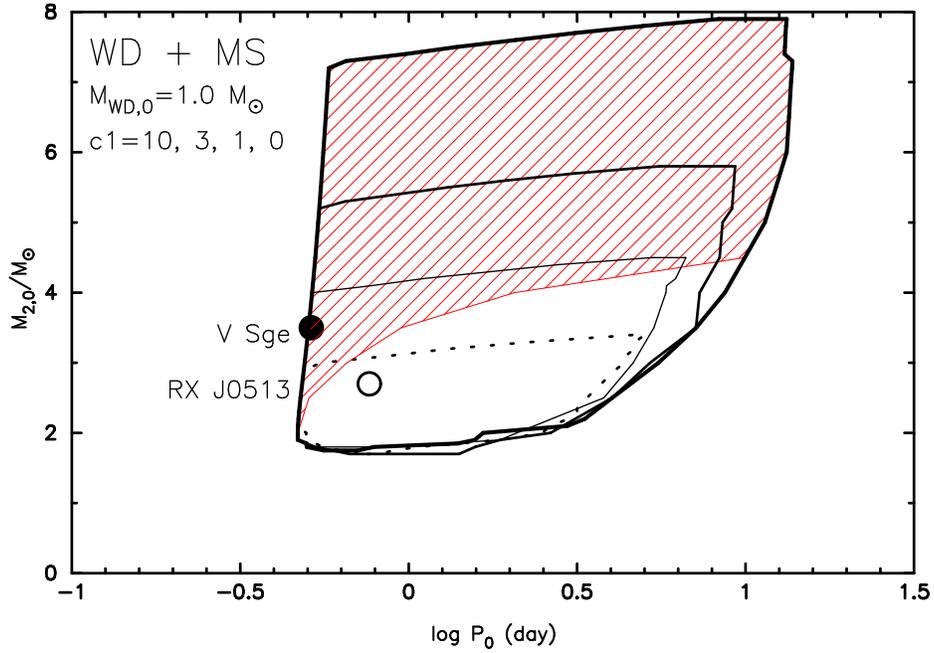}
\caption{
The initial parameter regions producing SNe Ia are plotted in
the $\log P_0 - M_{2,0}$ (orbital period --- donor mass) plane
for the WD + MS systems with various mass-stripping factors, $c_1$.
{\it Thick solid:} $c_1 = 10$.
{\it Medium solid:} $c_1 = 3$.  {\it Thin solid:} $c_1 = 1$.
{\it Dotted:} $c_1 = 0$.
The (red) hatched region indicates a region with a short delay time
($t_{\rm delay} \le 100$~Myr) for the case of $c_1 = 10$.
The region extends to the more massive donors for the larger $c_1$.
Two supersoft X-ray sources, RX~J0513.9$-6951$ ({\it open circle})
and V Sge ({\it filled circle}), are plotted, masses of which are
estimated to be $2.7 ~M_\sun$ \citep{hac03kb} and $3.5 ~M_\sun$ 
\citep{hac03kc}, orbital periods of which are determined to be
0.76 days \citep{pak93} and 0.51 days \citep{her65, pat98},
respectively.  The position of V~Sge suggests that $c_1 > 0$.
\label{zams_strip_reg100}}
\end{figure*}


\begin{figure*}
\epsscale{0.8}
\plotone{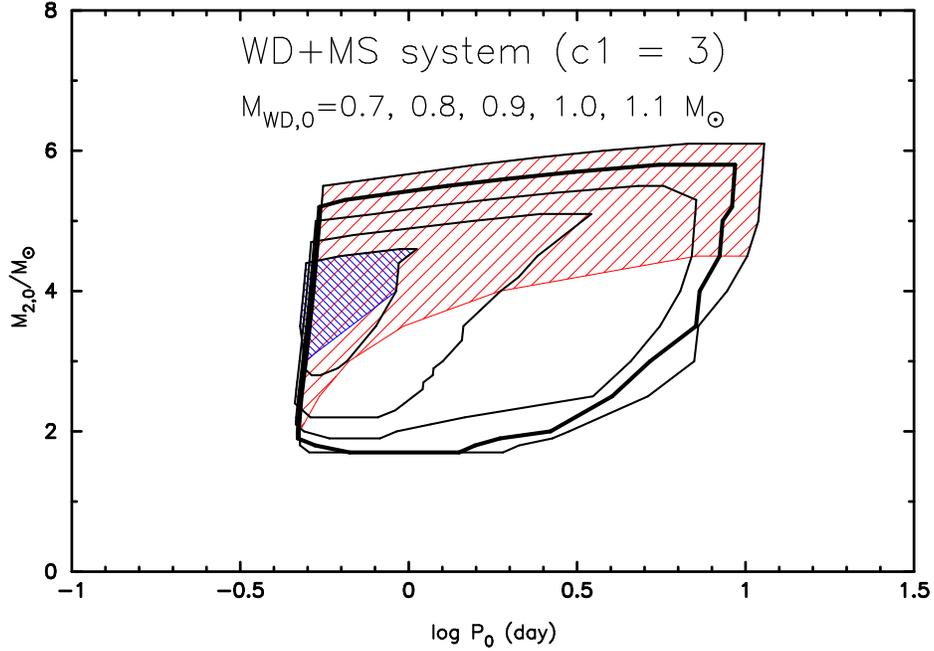}
\caption{
Dependence of the SN Ia parameter region on the initial WD mass,
$M_{\rm WD,0}$, for a mass-stripping factor of $c_1 = 3$.
From inside to outside, $M_{\rm WD,0} = 0.7$, 0.8, 0.9, 1.0
({\it thick solid line}), and $1.1 ~M_\sun$.
There is no region for $M_{\rm WD,0} = 0.6 ~M_\sun$.
The (red) sparse hatched region indicates the delay time of
$t_{\rm delay} \le 100$~Myr for $M_{\rm WD,0} = 1.1 ~M_\sun$ but
the (blue) dense hatched region for $M_{\rm WD,0} = 0.7 ~M_\sun$.
\label{zams_strip_reg_allmass}}
\end{figure*}


\begin{figure*}
\epsscale{0.8}
\plotone{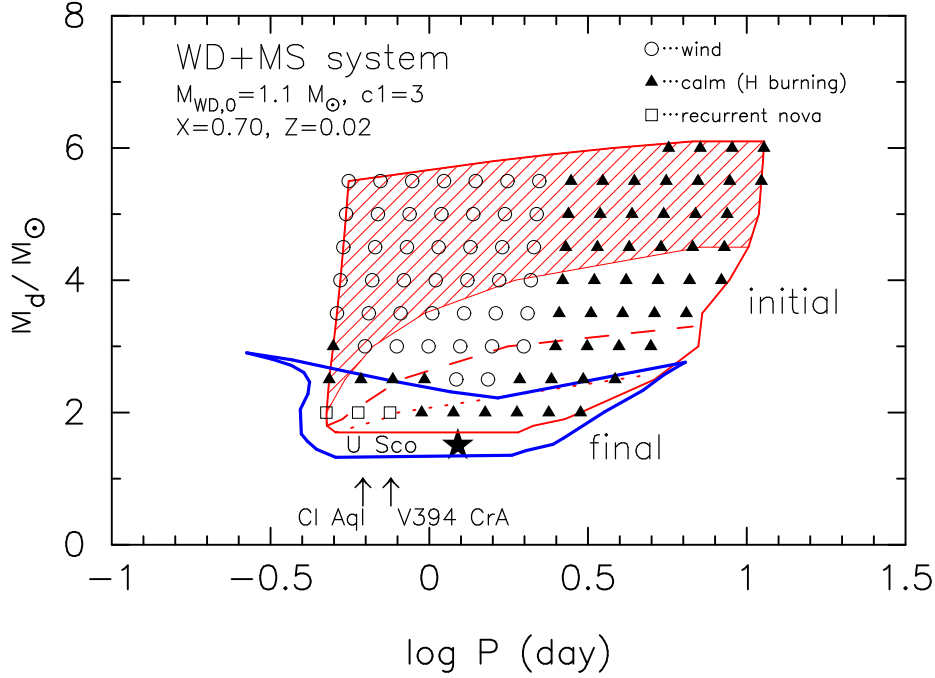}
\caption{
The parameter region that produces SNe Ia is plotted
in the $\log P - M_{\rm d}$ (orbital period --- donor mass) plane
for the WD + MS system.
Here we assume $M_{\rm WD,0}= 1.1 ~M_\sun$ for the initial white dwarf mass.
The initial WD + MS system inside the region encircled by the (red)
thin solid line (labeled ``initial'') will increase its white dwarf mass
up to the critical mass ($M_{\rm Ia}= 1.38 M_\sun$)
for the SN Ia explosion to occur. 
The final state of the WD + MS system in the $\log P - M_{\rm d}$ plane
just before the SN Ia explosion is encircled by the (blue)
thick solid line (labeled ``final'').
The final state of the WD just before the SN Ia explosion
is specified by one of wind ({\it open circle}),
steady H-burning ({\it filled triangle}),
or recurrent nova ({\it open square}) phase.
An hatched region indicates a region in which the progenitor
explodes in a delay time of $t_{\rm delay} \le 100$~Myr.
{\it Dashed line}: in a delay time of 200~Myr.
{\it Dotted line}: in a delay time of 400~Myr.
Currently known positions of three recurrent novae are indicated
by a star mark ($\star$) for U Sco \citep[e.g.,][]{sch95, hkkm00, hkkmn00},
and by arrows for the other two recurrent novae, V394 CrA
\citep{sch90} and CI Aql \citep{men95}, of unknown companion masses.
The WD masses of U~Sco and V394~CrA were estimated to be $1.37 ~M_\sun$
\citep{hkkm00, hac00kb} while that of CI~Aql was $1.2 ~M_\sun$
\citep{hac03ka}.
\label{zams_evl_con_c3_m110}}
\end{figure*}


\begin{figure*}
\epsscale{0.8}
\plotone{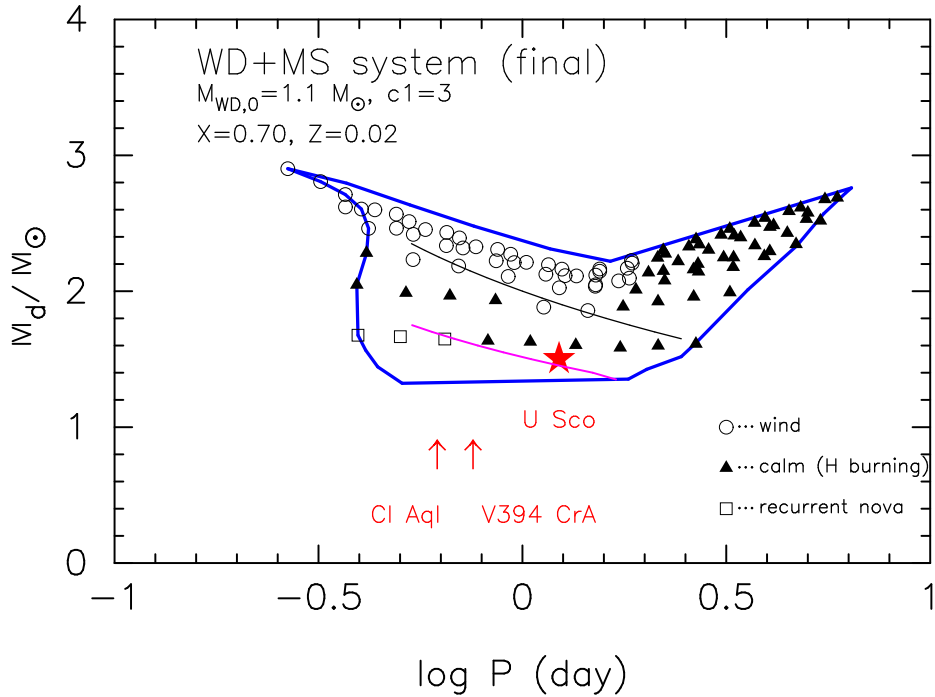}
\caption{
The final SN Ia region just before an SN Ia explosion.
Each symbol has the same meaning as 
in Fig. \ref{zams_evl_con_c3_m110}.
The upper black solid line and lower magenta solid line 
denote lines at $- {\dot M}_2 = {\dot M}_{\rm cr}$
and $- {\dot M}_2 = {\dot M}_{\rm stable}$, respectively,
just at the SN Ia explosion, where ${\dot M}_2$ is calculated from
eq. (\ref{m2dot}) with $R_2$ and $L_2$ taken from a single star
evolution given by \citet{tou97}.
Both the lines agree reasonably with the borders of
WIND--CALM and CALM--RN, respectively.
\label{zams_final_con_c3_m110}}
\end{figure*}


\begin{figure*}
\epsscale{0.8}
\plotone{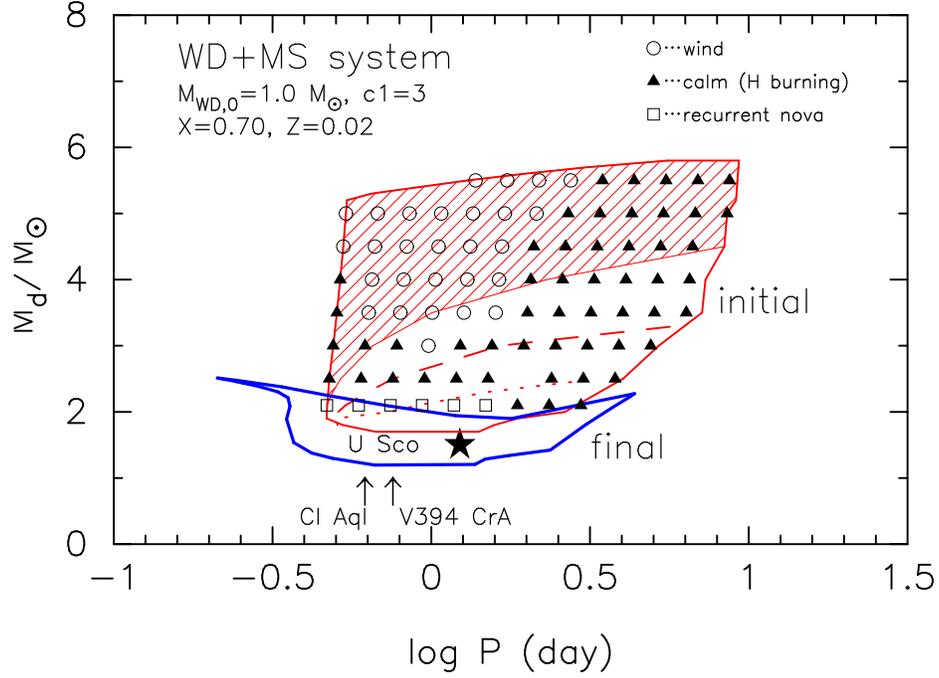}
\caption{
Same as Fig. \ref{zams_evl_con_c3_m110}, but for
an initial WD mass of $M_{\rm WD,0}= 1.0 ~M_\sun$.
\label{zams_evl_con_c3_m100}}
\end{figure*}


\begin{figure*}
\epsscale{1.1}
\epsscale{0.8}
\plotone{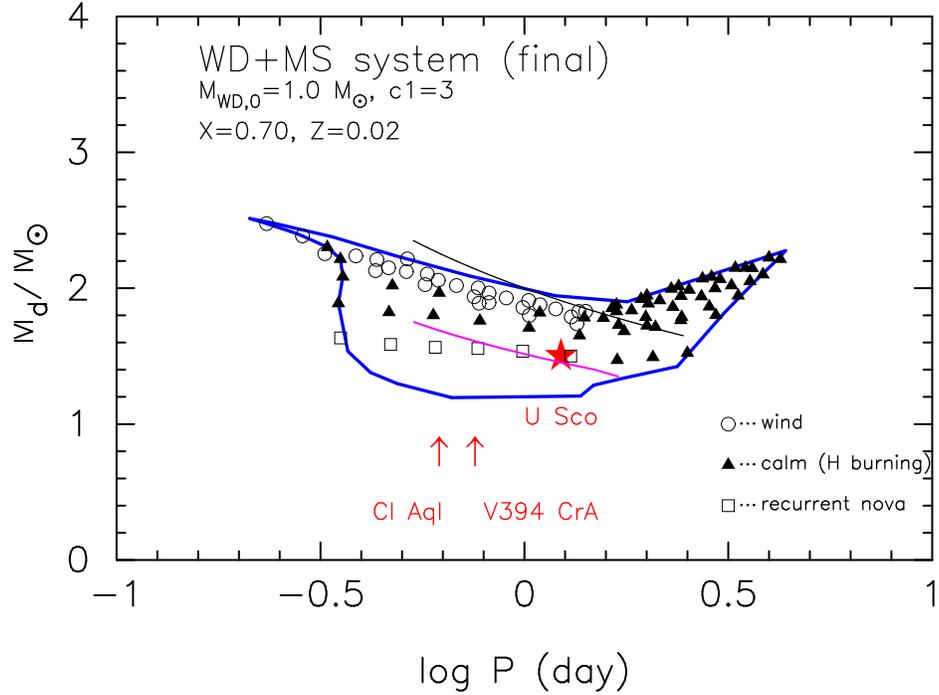}
\caption{
Same as Fig. \ref{zams_final_con_c3_m110}, but for
an initial white dwarf mass of $M_{\rm WD,0}= 1.0 ~M_\sun$.
Large difference in the border of WIND--CALM comes from 
the fact that the secondary considerably overfills the Roche lobe,
i.e., $R_2 > R_2^*$, at the SN Ia explosion in the Case WIND. 
\label{zams_final_con_c3_m100}}
\end{figure*}


\begin{figure*}
\epsscale{0.8}
\plotone{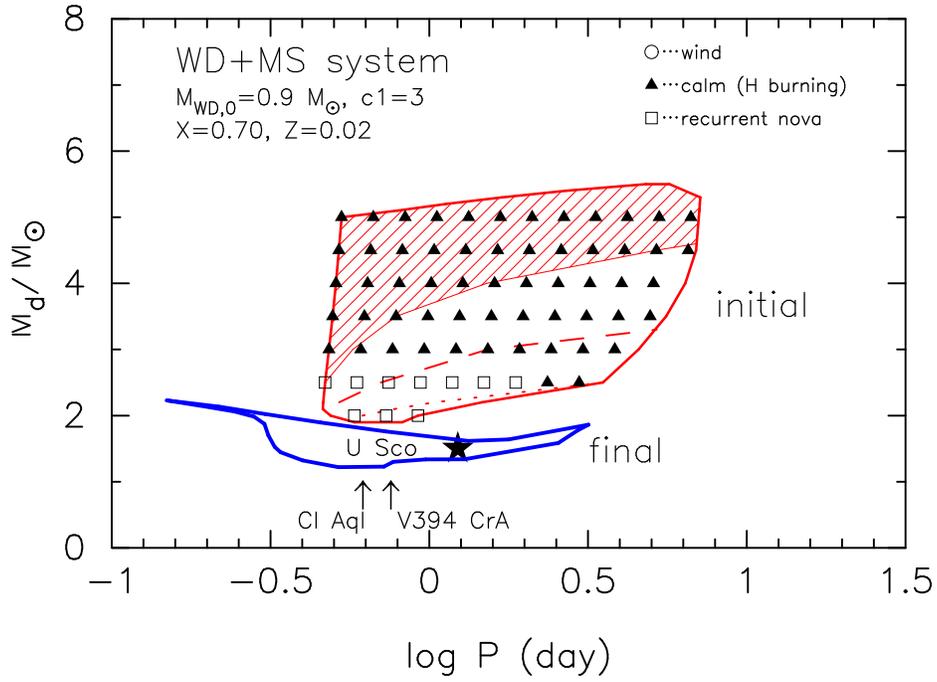}
\caption{
Same as Fig. \ref{zams_evl_con_c3_m110}, but for
an initial white dwarf mass of $M_{\rm WD,0}= 0.9 ~M_\sun$.
There is no Case WIND (no {\it open circles}).
\label{zams_evl_con_c3_m090}}
\end{figure*}


\begin{figure*}
\epsscale{0.8}
\plotone{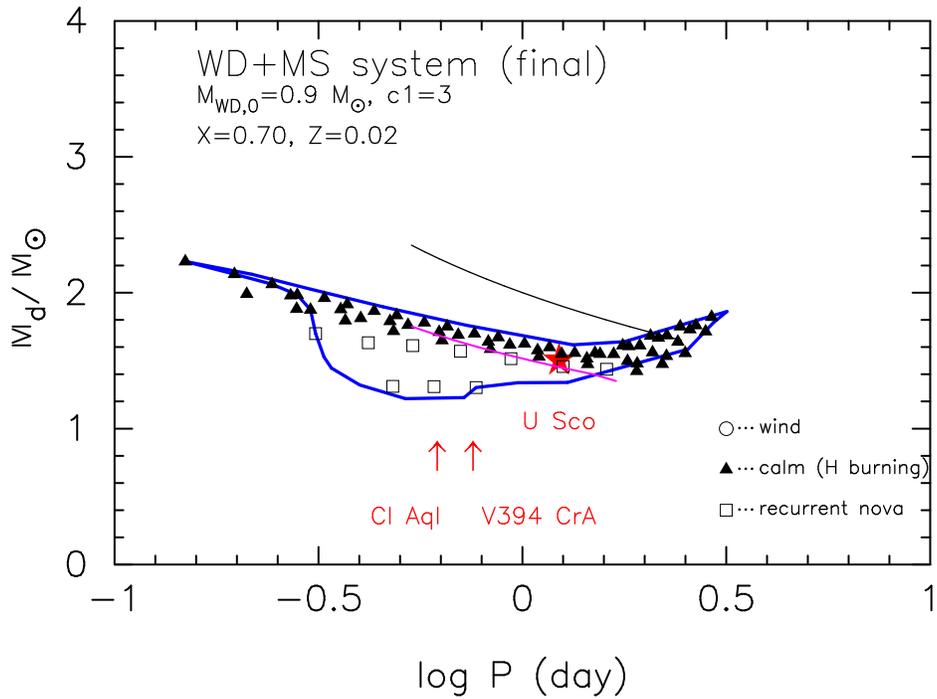}
\caption{
Same as Fig. \ref{zams_final_con_c3_m110}, but for
an initial white dwarf mass of $M_{\rm WD,0}= 0.9 ~M_\sun$.
\label{zams_final_con_c3_m090}}
\end{figure*}


\begin{deluxetable*}{llllllll}
\footnotesize
\tablecaption{Three typical cases of SN Ia explosion
\label{sn1a_explosion}}
\tablewidth{0pt}
\tablehead{
\colhead{case} & \colhead{wind} & \colhead{H burning}
& \colhead{CSM} & \colhead{pre-SN history}
& \colhead{SN Ia}  & \colhead{delay time} & \colhead{immediate radio/X-ray}
}
\startdata
WIND & wind & steady & massive: near & WIND (V~Sge type) & IIa (02ic-like) & young & yes \nl
CALM & no wind & steady & thin: far & WIND$\rightarrow$SSXS & normal Ia & young & no ($\sim 10-100$ yr) \nl
RN & no wind & flash & very thin: many shells & WIND$\rightarrow$SSXS$\rightarrow$RN & normal Ia & broad & no ($\sim 100$ -- 1000 yr) \\
  &  &  &  & or SSXS$\rightarrow$RN   &  &  &
\enddata
\end{deluxetable*}

\section{Young Population Type Ia Supernovae}
Based on the binary evolution scenario proposed by \citet{hkn99, hknu99},
we have followed binary evolutions starting from stage (b) in
Figure \ref{stripping_evolution}, that is, just when the companion
evolves to fill its Roche lobe.  The main difference from the previous
work cited above is the inclusion of mass-stripping effect.
Our results are shown in Figures
\ref{zams_strip_reg100}--\ref{zams_final_con_c3_m090}.

Figure \ref{zams_strip_reg100} shows the parameter regions that produce
SNe Ia ({\sl SN Ia region}) in the $\log P_0 - M_{2,0}$ (the initial
orbital period and the initial secondary mass) plane for the WD + MS
system.  Here the initial white dwarf mass is assumed to be $M_{\rm
WD,0}= 1.0 ~M_\sun$.  The white dwarfs within these SN Ia regions will
increase their mass, $M_{\rm WD}$, up to the critical mass
($M_{\rm Ia}= 1.38 ~M_\sun$) for the SN Ia explosion to occur. 

The SN Ia region in the $\log P_0 - M_{2,0}$ plane is enclosed by four
boundaries.  (1) The left boundary is given by the mass-radius
relation for the zero-age main-sequence stars.  (2) The lower boundary
is set by strong nova explosions, below which 
$\dot M_{\rm transfer}\lesssim 1 \times 10^{-7} M_\sun$~yr$^{-1}$ and the
resultant nova explosion ejects most of the accreted matter, thus
preventing the WD mass from increasing.  (3) The upper boundary is
limited by the formation of a common envelope.  Here we assume that
a common envelope is formed when $\dot M_{\rm transfer}
\gtrsim 1 \times 10^{-4} M_\odot$ yr$^{-1}$ because
$R_{\rm 1,ph} \gtrsim a \sim 10 ~R_\odot$
for such a high $\dot M_{\rm transfer}$
\citep[see][for more details]{hknu99}.  (4) The right boundary
corresponds to the end of central hydrogen burning
of the MS companion: after that, it shrinks and underfills its Roche lobe.

In Figure \ref{zams_strip_reg100}, the SN Ia regions for the various
mass-stripping factor $c_1 =$ 10, 3, and 1 are encircled by the thick,
medium, and thin solid lines, respectively, and the no stripping case
($c_1 = 0$) by the dotted line.  The position of the Galactic
supersoft X-ray source V~Sge is clearly outside the SN Ia region
for $c_1 =0$, but inside the SN Ia region if $c_1 > 0$.
For larger $c_1$, the SN Ia region extends to more massive
$M_{2,0}$, because the stronger mass-stripping leads to the lower mass
transfer rate, $\dot M_{\rm transfer}$, from the secondary to the WD
(see eq.[\ref{net_mass_transfer_on_c1}]),
thus preventing the formation of a common envelope for larger $M_{2,0}$.
As shown in this figure quite massive secondaries produce SNe Ia
(e.g., $M_{2,0} = 7.5 ~M_\sun$ for $c_1 = 10$)
for the strong mass-stripping case of $c_1 \gtrsim 3$.

Such WD + MS systems with a massive MS secondary consist of a very young
population of SNe Ia.  We show the region of short delay times,
$t_{\rm delay} \le 100$~Myr, by the red shadow
in Figures \ref{zams_strip_reg100}, \ref{zams_strip_reg_allmass},
\ref{zams_evl_con_c3_m110}, \ref{zams_evl_con_c3_m100}, and
\ref{zams_evl_con_c3_m090}.  Figure
\ref{zams_strip_reg_allmass} shows the SN Ia regions for different
initial WD masses, $M_{\rm WD,0}= 0.7$, 0.8, 0.9, 1.0, and $1.1
~M_\sun$.  The red (sparse) and blue (dense) hatched regions
indicate the delay time of $t_{\rm delay} \le
100$~Myr for $M_{\rm WD,0} = 1.1 ~M_\sun$ and $0.7 ~M_\sun$, respectively.

We apply the present result to equation (1)
of \citet{ibe84}, i.e.,
\begin{equation}
\nu = 0.2 \cdot \Delta q \cdot
\int_{M_l}^{M_u} {{d M} \over M^{2.5}} \cdot \Delta \log ~a
\quad {\rm yr}^{-1},
\label{realization_frequency}
\end{equation}
where $\Delta q$, $\Delta \log ~a$, $M_l$, and $M_u$ are
the appropriate ranges of the mass ratio and the initial separation,
and the lower and upper limits of the primary mass
for SN Ia explosions in solar mass units, respectively.
We then estimate the SN Ia birth rate in our Galaxy as $\nu_{\rm WD+MS}
\sim 0.004$~yr$^{-1}$, which is consistent with the observation
\citep{cap99}.

On the other hand, \citet{hkn99} proposed another channel to
SNe Ia, the symbiotic channel, binary of which consists of
a white dwarf and a red giant (WD + RG), and estimated its birth
rate to be $\nu_{\rm WD+RG} \sim 0.002$~yr$^{-1}$.

Assuming the initial distribution of binaries given by
equation (\ref{realization_frequency}) at the burst of
star formation (single event),
we estimate the delay time distribution of SNe Ia
for the WD + MS systems in Figure \ref{birth_rate_population_ms}.
The number ratio of these young populations is calculated for
10 bins of delay time, $(0.025, 0.05)$, $(0.05, 0.1)$, $(0.1, 0.2)$,
$(0.2, 0.4)$, $(0.4, 0.8)$, $(0.8, 1.6)$, $(1.6, 3.2)$, $(3.2, 6.4)$, 
$(6.4, 12.8)$, and $(12.8, 25.6)$ Gyr.
  The number ratio with $t_{\rm delay} \le 100$~Myr
and $t_{\rm delay} \le 200$~Myr are about 50\% and 80\%, respectively,
of the total SNe Ia coming from the
WD + MS system, which is consistent with the recent observational
suggestions \citep[e.g.,][]{man06, aub07}.

Short delay times ($t_{\rm delay} \lesssim 10^8$~yr) of some SNe Ia
have been suggested from the distribution of
SNe Ia relative to spiral arms \citep[e.g.,][]{bar94, del94}.
\citet{pet05} reported that about 30--40\% of SNe Ia are associated
with spiral arms in their samples, being consisting with our results.
\citet{man06} have suggested that the delay time distribution
function of SNe Ia has a bimodality, one for young
population ($t_{\rm delay} \sim 100$~Myr) and the other with
a broad distribution over $\sim 3$ Gyr.
Our delay time distribution function has a peak around
$t_{\rm delay} \le 100$ Myr from the WD + MS systems and a broad
distribution from the WD + RG systems \citep{hkn99} as shown in
Figure \ref{birth_rate_population_ms_rg}.

\begin{deluxetable*}{llllll}
\footnotesize
\tablecaption{Initial parameters for Three SN Ia explosions
\label{condition_sn1a_explosion}}
\tablewidth{0pt}
\tablehead{
\colhead{WD mass} & \colhead{secondary mass} & \colhead{orbital period}
& \colhead{case} & \colhead{pre-SN history} & \colhead{SN Ia}
\nl
\colhead{($M_\sun$)} & \colhead{($M_\sun$)} & \colhead{(days)}
& \colhead{} & \colhead{} & \colhead{}
}
\startdata
$1.0-1.1$ & $3-6$ & $\sim 0.5-2$ & WIND & WIND & IIa (02ic-like) \nl
$1.0-1.1$ & $3-6$ & $\sim 2-10$ & CALM & WIND$\rightarrow$SSXS & normal Ia \nl
$1.0-1.1$ & $2.2-3$ & $\sim 0.5-4$ & CALM & WIND$\rightarrow$SSXS & normal Ia \nl
$1.0-1.1$ & $1.8-2.2$ & $\sim 0.5-2$ & RN & WIND$\rightarrow$SSXS$\rightarrow$RN
& normal Ia \nl
 &  &  & & or SSXS$\rightarrow$RN & \nl
$0.9$ & $2.5-5$ & $\sim 0.5-6$ & CALM & WIND$\rightarrow$SSXS
& normal Ia \nl
$0.9$ & $2.0-2.5$ & $\sim 0.5-2$ & RN & WIND$\rightarrow$SSXS$\rightarrow$RN
& normal Ia \nl
$0.8$ & $4-5$ & $\sim 1-3$ & RN & WIND$\rightarrow$SSXS$\rightarrow$RN
& normal Ia \nl
$0.8$ & $4-5$ & $\sim 0.5-1$ & CALM & WIND$\rightarrow$SSXS
& normal Ia \nl
$0.8$ & $2.5-4$ & $\sim 0.5-2$ & RN & WIND$\rightarrow$SSXS$\rightarrow$RN
& normal Ia \nl
$0.7$ & $3-4.5$ & $\sim 0.5-1$ & RN & WIND$\rightarrow$SSXS$\rightarrow$RN
& normal Ia 
\enddata
\end{deluxetable*}

\section{Final Stage of Binary Evolution and Circumstellar Matter}
The final state of the WD depends mainly on the mass transfer rate
${\dot M}_{\rm transfer}$ from the donor star to the WD at the SN Ia
explosion \citep{nom82, hkn99, nom07}.
As shown in Figure \ref{evolution_sn2005gj}, $\dot M_2$ drops 
quickly in the early stage and then slows down to almost a
constant value.  At least, in the early phase, the mass transfer
proceeds on a thermal time scale, represented by the second term
of equation (\ref{secondary_mass_transfer}),
when the mass ratio $M_2/M_1$ exceeds 0.79.  So we approximate
the mass transfer rate as
\begin{equation}
- {\dot M}_2 \approx {{M_2} \over{ \tau_{\rm KH}}} \sim 
3 \times 10^{-8} ~M_\sun {\rm ~yr~}^{-1} \left({{R_2} \over {R_\sun}} \cdot 
{{L_2} \over {L_\sun}}\right)\left({{M_2} \over {M_\sun}}
\right)^{-1} 
\label{m2dot}
\end{equation}
By applying the approximate $M_2 - L_2$ relation of $L_2 \propto M_2^m$,
where $m \sim 4$ for the $1.5 - 3 ~M_\sun$ zero-age main-sequence (ZAMS) 
stars or $m \sim 3.5$ for the $3- 7~M_\sun$ ZAMS stars,
\begin{equation}
   - {\dot M}_2 \propto R_2 M_2^{m-1}.
\label{m2dot_approx}
\end{equation}
Thus $- {\dot M}_2$ decrease as $M_2$ decreases.

Figures \ref{zams_evl_con_c3_m110}--\ref{zams_final_con_c3_m090}
show the SN Ia regions in the $\log
P - M_{\rm d}$ (orbital period --- donor mass) plane for the
initial WD + MS system (encircled by the red thin line and labeled
``initial'') as well as the final state at the SN Ia explosion
(encircled by the blue thick line and labeled ``final'').
Here we assume $c_1 = 3$ and $M_{\rm WD,0}= 1.1$, 1.0,
and $0.9 ~M_\sun$.  In these figures, we distinguish
three final states just before the SN Ia explosion, i.e.,
optically thick WD wind phase (WIND: open circles), steady hydrogen
burning phase without optically thick winds from WDs 
(CALM: filled triangles), and recurrent nova (RN) phase
(RN: open squares).  The characteristic properties for these three
progenitor stages 
are summarized in Table \ref{sn1a_explosion} and the corresponding 
binary parameters are tabulated in Table \ref{condition_sn1a_explosion}.

\subsection{Case WIND}
When the mass transfer rate from the secondary continuously exceeds
the critical rate of equation (\ref{critical_rate_wind}) until
the final stage, the WDs explode during the wind phase
(Fig. \ref{evolution_sn2005gj}a).  Therefore, we call this Case WIND.
Case WIND is realized in the region of $M_{2,0} \gtrsim 3
~M_\sun$ and $P_{2,0} \lesssim$ 2 days for $M_{\rm WD,0}= 1.1 ~M_\sun$
and $1.0 ~M_\sun$ (open circle), but no Case WIND exists for
$M_{\rm WD,0} \le 0.9 ~M_\sun$ as shown in Figures
\ref{zams_evl_con_c3_m110}, \ref{zams_evl_con_c3_m100},
and \ref{zams_evl_con_c3_m090}.

The stripped-off matter from the companion can easily amount to
$\Delta M_{\rm strip} \sim 1-2 ~M_\sun$ and even reach $3-4 ~M_\sun$ as
seen from the donor mass difference $\Delta M_2$ between the
``initial'' and the ``final'' in Figures
\ref{zams_evl_con_c3_m110}, \ref{zams_evl_con_c3_m100}, and
\ref{zams_evl_con_c3_m090}.
More precisely, $\Delta M_2$ consists of three parts,
the stripped-off mass $\Delta M_{\rm strip}$,
the accreted mass by the WD $\Delta M_1$,
and the mass ejected by the WD wind $\Delta M_{\rm wind}$, i.e.,
${\dot M}_2  = {\dot M}_{\rm strip} + {\dot M}_{\rm wind} - {\dot M}_1$
from equation (\ref{total_mass_conservation}).  This can be approximated
as $ \dot M_2 \approx \dot M_{\rm strip} + {\dot M}_{\rm wind} 
= (1+1/c_1) {\dot M}_{\rm strip} = 4/3 {\dot M}_{\rm strip}$ because
${\dot M}_1 \ll - {\dot M}_2$, so that
$\Delta M_{\rm strip} \approx 3/4 \Delta M_2$ for $c_1 = 3$.

The stripped-off material forms CSM very near the SN Ia. 
We expect that stripped-off matter did not go away from
the system because the velocity of stripped-off matter
may not exceed the escape velocity
of the binary system.  Then the SN Ia undergoes circumstellar
interaction as observed in Type Ia/IIn (or IIa) SNe 2002ic and 2005gj.

\citet{ald06} suggested that the host galaxy of SN 2005gj
had a burst of star formation $200 \pm 70$~Myr ago.
If the progenitor of SN 2005gj was born at that time,
its delay time is consistent with our Case WIND as shown
in Figure \ref{birth_rate_population_ms}.

\subsection{Case CALM}
When the mass transfer rate from the secondary is below the critical
rate for optically thick winds but above the lowest rate of
steady hydrogen burning, i.e.,
${\dot M}_{\rm stable} < {\dot M}_{\rm transfer} < {\dot M}_{\rm cr}$,
the WDs undergo steady H-burning at the time of SN Ia explosion
(filled triangles in
Figs. \ref{zams_evl_con_c3_m110}--\ref{zams_final_con_c3_m090}).
We call this Case CALM because no optically thick winds occur.
The WDs are observed as supersoft X-ray sources
(SSXSs) until the SN Ia explosion.
The stripped-off material forms CSM but it has been dispersed
too far to be detected immediately after the SN Ia explosion.  

The CALM case is realized 
in the region of $M_{2,0} \gtrsim 3 ~M_\sun$ and $P_{2,0} \gtrsim$ 2 days
for $M_{\rm WD,0}= 1.1 ~M_\sun$ and $1.0 ~M_\sun$ in Figures
\ref{zams_evl_con_c3_m110}, \ref{zams_evl_con_c3_m100}, and
\ref{zams_evl_con_c3_m090}, where ${\dot M}_{\rm transfer}$
in the early phase is much larger than that of
$P_{2,0} \lesssim$ 2 days, because in equations 
(\ref{secondary_mass_transfer}) and (\ref{m2dot}),
$R_2$ and $L_2$ are much larger than those for $P_{2,0} \lesssim$ 2 days. 
Then $\dot M_{\rm transfer}$ is much larger,
thus much more mass had been lost in the earlier phase.
As a result, the wind phase finishes at an earlier time
even for the same initial mass $M_{2,0}$ as seen in
Figures \ref{evolution_sn2005gj}a and \ref{evolution_sn2005gj}b.
Therefore, at the SN Ia explosion, no wind occurs.

In the region of $M_{2,0} \lesssim 3 ~M_\sun$ 
for $M_{\rm WD,0}= 1.1 ~M_\sun$ and $1.0 ~M_\sun$ 
in Figures \ref{zams_evl_con_c3_m110}--\ref{zams_final_con_c3_m100}
(filled triangles or open squares),
$M_2$ decreases to as small as the primary
$M_{\rm WD}$, i.e., the mass ratio of $q \sim 1$, at the SN Ia explosion,
which corresponds to a lower part of the ``final'' region.
Then the mass transfer rate decreases down to
$\dot M_{\rm transfer} < \dot M_{\rm cr}$
or even $\dot M_{\rm transfer} < \dot M_{\rm stable}$
because $L_2$ is smaller for the smaller
$M_2$ even if the mass transfer itself is proceeding
on a thermal time scale.  
The wind phase has ended earlier than for $M_{2,0} \gtrsim 3~M_\sun$.

The border between Case WIND and Case CALM can be simply estimated
from the condition ${\dot M}_{\rm transfer} = {\dot M}_{\rm cr}$ and
$R_2 = R_2^*$ at the SN Ia explosion.  We have calculated ${\dot M}_2$
from equation (\ref{m2dot}) with $R_2$ and $L_2$ being taken from
\citet{tou97} (a single star evolution).
This line agrees reasonably with the border of WIND--CALM
in Figure \ref{zams_final_con_c3_m110} but largely deviates
from it in Figure \ref{zams_final_con_c3_m100}.
This is because the secondary considerably overfills the Roche lobe,
i.e., $R_2 > R_2^*$, at the SN Ia explosion for
$M_{\rm WD,0}= 1.0 ~M_\sun$. 

For $M_{\rm WD,0} \lesssim 0.9 ~M_\sun$ (filled triangles
or open squares), $M_2$ decreases 
at the SN Ia explosion as shown in Figure \ref{zams_evl_con_c3_m090}
(``final'' region) and ${\dot M}_{\rm transfer}$ decreases to be lower
than ${\dot M}_{\rm cr}$ mainly 
because the time for the WD to reach $M_{\rm Ia}= 1.38 ~M_\sun$ is
longer and much more mass is lost during the evolution.
The wind phase has ended before the SN Ia explosion.

For a typical case of $c_1=3$, $M_{\rm WD,0}= 0.9 ~M_\sun$,
$M_{2,0}= 4.0 ~M_\sun$, and $P_0 = 1.3$ days
in Figure \ref{zams_evl_con_c3_m090}, the WD explodes as an SN Ia
at $t= 9 \times 10^5$~yr after the secondary fills its Roche lobe.  
The wind has already stopped $3 \times 10^5$~yr ago
(duration of the WIND phase, $\Delta t_{\rm wind}= 6 \times 10^5$ yr,
and duration of the CALM phase, $\Delta t_{\rm calm}= 3 \times 10^5$ yr),
so that the inner edge of stripped-off material has already gone
to ($10-100$ km~s$^{-1})~ \times (3 \times 10^5$~yr) $\sim 10^{19}
- 10^{20}$~cm from the SN Ia.
Therefore, it takes about $10-100$ yr for the SN Ia
ejecta to reach the inner edge of stripped-off matter.
We do not expect radio or X-ray until,
at least, $10-100$ yr after the explosion.
Thus the resultant SNe Ia are mostly ``normal.''  The duration of
CALM phase is typically a third or fourth of the total
evolution time to the SN Ia.  These long durations of optically
thick wind phases may reduce the statistical number of luminous supersoft
X-ray sources because the photospheric temperature of the WD is
lower than $\sim 10$~eV and not luminous in supersoft X-ray.

The decline of SN 2006X light curves is slowing down
in a later phase compared with the other normal SNe Ia light curves,
suggesting an interaction between the ejecta and CSM in a later phase
\citep{wan07a} or a light echo of circumstellar/interstellar
matter \citep{wan07b}.
This happens if SN 2006X is placed at the border between
our Case WIND and Case CALM since the innermost part of
slowly expanding circumstellar matter has not yet moved far away.
A X-ray detected SN 2005ke may also belong to the same category
\citep{imm06}.

For the progenitor of SN 2006X, 
\citet{pat07a} suggested a WD + RG system like RS~Oph from
the circumstellar matter (CSM) absorption lines.  Here we suggest that
a WD + MS system (like U~Sco) may better explain a continuous velocity
distribution (from $\sim - 30$ to $- 150$ km~s$^{-1}$)
of the CSM absorption lines by the stripped
matter with continuous velocity distribution
(see Fig. \ref{stripping_evolution}d).  In this connection,
very recent report of the \ion{Na}{1}~D circumstellar lines of RS~Oph
during the 2006 outburst is suggestive \citep{iij07}.
These lines indicate no continuous distribution
as observed in SN 2006X but a narrow velocity component of $-36$~km~s$^{-1}$
against RS~Oph that is attributed to the red giant cool wind.

Recently negative detections of time-variable \ion{Na}{1} D lines
have been reported for two SNe Ia 2000cx \citep{pat07b} and 2007af
\citep{sim07}.  \citet{pat07b} and \citet{sim07} suggested a
possibility that the distribution of CSM is torus/disk-like 
as illustrated in Figure \ref{stripping_evolution}.  In such a case,
variable \ion{Na}{1} D lines would not be observed if the
line of sight is perpendicular to or off the orbital plane.
Since the hot WD winds have a large velocity of
$\gtrsim 1000$~km~s$^{-1}$, the CSM formed by hot winds
quickly diffuse away and is too tenuous to be detected.

Recently, \citet{bad07} reported that the fast WD wind of
$v \gtrsim 200$~km~s$^{-1}$, which excavates its circumstellar medium
and forms a large cavity around an SN Ia, is incompatible
with the X-ray emission from the shocked ejecta in our Galaxy
(Kepler, Tycho, SN 1006), Large Magellanic Cloud (0509-67.5, 0519-69.0,
N103B), and M31 (SN 1885).  We can avoid this difficulty
if the stripped-off matter has a velocity of $10-100$~km~s$^{-1}$.

\subsection{Case RN}
When the mass transfer rate from the secondary is below
the lowest rate of steady hydrogen burning, i.e.,
${\dot M}_{\rm transfer} < {\dot M}_{\rm stable}$, hydrogen shell burning
is unstable to flash and recur many times in a short period as a
recurrent nova (RN) (the open squares in Figs.
\ref{zams_evl_con_c3_m110}--\ref{zams_final_con_c3_m090}).
We call this Case RN.  
The recurrent nova U~Sco,
one of the candidates of SN Ia progenitors, is in the middle of the
``final'' region \citep{hkkm00, hkkmn00}.
The resultant explosions are ``normal'' SNe Ia.

A simple estimation gives the border between Case CALM and Case RN,
${\dot M}_{\rm transfer}  = {\dot M}_{\rm stable}$ 
and $R_2 = R_2^*$ at the SN Ia explosion.
Here we calculate ${\dot M}_2$ from equation (\ref{m2dot})
with $R_2$ and $L_2$ being taken from \citet{tou97}
(a single star evolution).  These lines agree reasonably
with the border of CALM--RN in Figures 
\ref{zams_final_con_c3_m110}, \ref{zams_final_con_c3_m100},
and \ref{zams_final_con_c3_m090}.

For a typical Case RN of $M_{\rm WD,0}=1.0 ~M_\sun$, 
$M_{\rm 2,0}=2.0 ~M_\sun$, $P_0=1.18$ days with $c_1=3$, 
the WD undergoes the SN Ia explosion in the recurrent nova phase
at $t= 9.49 \times 10^5$ yr after the secondary first fills its Roche
lobe.  The WD wind stops at $t = 4 \times 10^5$~yr
and the stable hydrogen burning ends at $t= 8.6 \times 10^5$~yr.
During the last $10^5$~yr in the recurrent nova phase,
the secondary loses $\sim 0.022~M_\sun$, of which the WD accretes
$0.017~M_\sun$.  Therefore the stripped-off matter
in the recurrent nova phase is very small.
On the other hand,  the stripped-off matter in the
early wind phase amounts $\Delta M_{\rm strip} \approx 0.15~M_\sun$,
which has already been far from the SN at the SN Ia
explosion, i.e., (10--100 km~s$^{-1})~ \times ~(5 \times 10^5$~yr) $=
(1-10)\times 10^{19}$~cm.   It takes about 100-1000 yr for the SN ejecta
to reach the stripped-off matter.  These features are summarized in
Table \ref{sn1a_explosion}.

\begin{figure*}
\epsscale{0.8}
\plotone{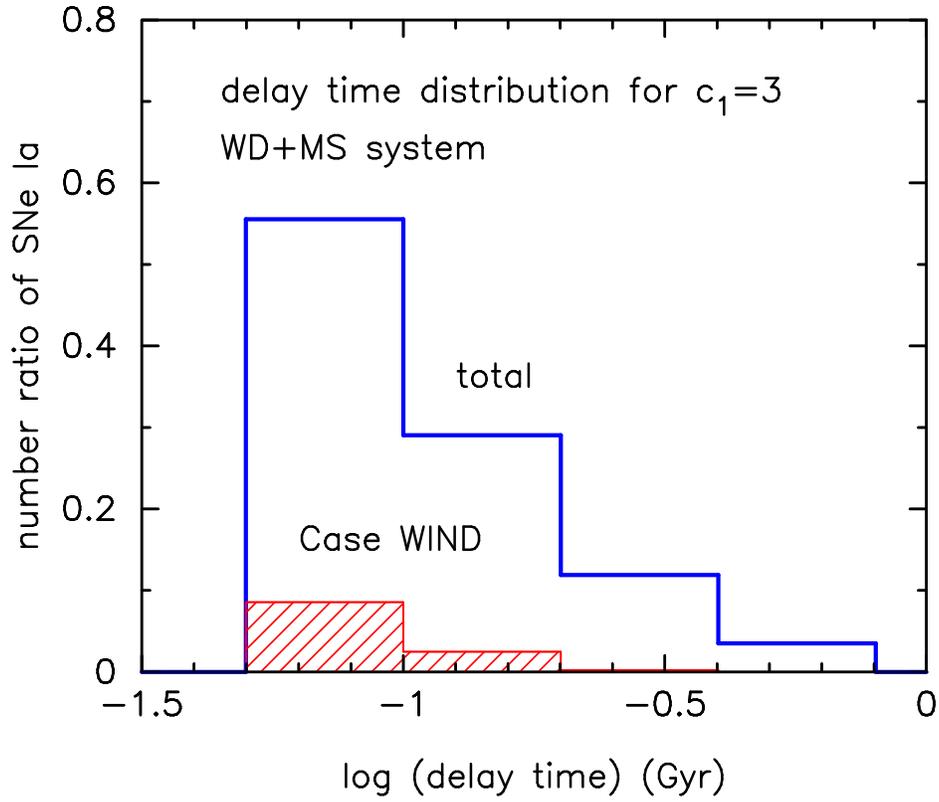}
\caption{
Upper blue thick histogram: delay time distribution for $c_1 = 3$.
Each bin is separated
by the delay time $(0.05, ~0.1)$, $(0.1, ~0.2)$, $(0.2, ~0.4)$, 
and $(0.4, ~0.8)$ Gyr.  About 50\% of SNe Ia coming
from the WD + MS systems explode in within 0.1 Gyr.
Lower red shadowed histogram: the ratio of SN 2002ic type
(Case WIND) SNe Ia.
About 7\% of SNe Ia coming from the WD + MS systems explode
in a wind phase.
\label{birth_rate_population_ms}}
\end{figure*}

\begin{figure*}
\epsscale{0.8}
\plotone{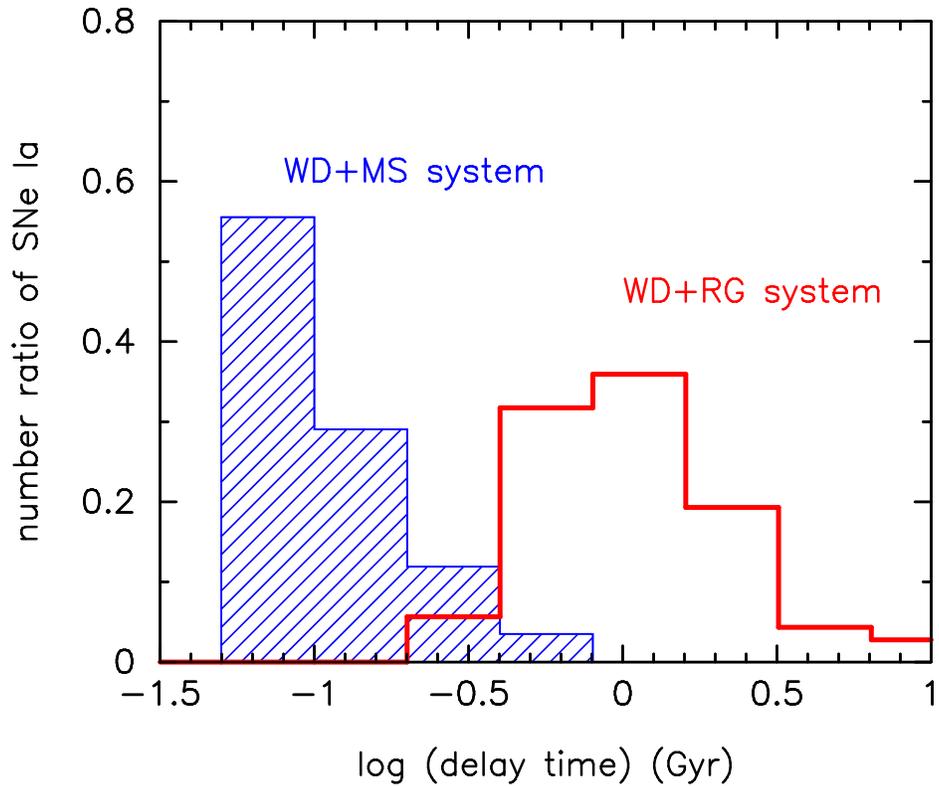}
\caption{
Same as Fig. \ref{birth_rate_population_ms}, but for both
the WD + MS (blue shadowed histogram) and WD + RG (red thick
histogram) systems.  Each bin is separated
by the delay time $(0.05, ~0.1)$, $(0.1, ~0.2)$, $(0.2, ~0.4)$, 
$(0.4, ~0.8)$, $(0.8, ~1.6)$, $(1.6, ~3.2)$, $(3.2, ~6.4)$, 
and $(6.4, ~12.8)$ Gyr.  The number ratio is normalized for
each system. 
\label{birth_rate_population_ms_rg}}
\end{figure*}

\section{Discussion}
\subsection{Mass-stripping Effect and Modulated Mass Transfer Rate}
As mentioned in \S 1, the existence of mass-stripping effect has been
demonstrated by \citet{hac03kb, hac03kc}.  They analyzed two quasi-periodic
transient supersoft X-ray sources, RX~J$0513.9-6951$ in the
Large Magellanic Cloud (LMC) and V~Sge in our Galaxy.
Especially V~Sge shows the following key observational features:
(1) V~Sge exhibits long-term transitions between
optical high (its brightness of $V \sim 11$ and its duration of 
$\sim 180$ days) and low ($V \sim 12$ and $\sim 120$ days) states
with total durations of $\sim 300$~days
\citep[see, e.g.,][for the long-term behavior]{sim99}.
(2) Very soft but very weak X-rays are detected only in the long-term
optical low state \citep[e.g.,][]{gre98}.
(3) Radio observations indicate a wind mass-loss rate as large as
$\sim 10^{-5}M_\sun$~yr$^{-1}$ \citep*{loc97, loc99}.

\citet{hac03kc} explained these features based on the mass-stripping
effect; the mass transfer to the WD is modulated by the WD wind because
mass-stripping attenuates the mass transfer rate.  This interaction
leads to high and low states.  The mass loss rate of the WD wind 
(with a high velocity of $\gtrsim 1000$ km~s$^{-1}$) reaches
as high as ${\dot M}_{\rm wind} \sim 1 \times 10^{-5} M_\sun$~yr$^{-1}$,
being consistent with the radio observation.  Thus, the mass transfer
rate itself may not be constant but vary in time, thus being regarded
as a time-averaged rate in the present paper.

From the light-curve fitting Hachisu \& Kato also estimated the WD
mass as $M_{\rm WD} \sim 1.25 ~M_\sun$ and the secondary mass to be
$M_{\rm MS} \sim 3.5 ~M_\sun$, and
concluded that V~Sge will explode as an SN Ia
in a time scale of $\sim 1 \times 10^5$~yr.
Since the present orbital period of V~Sge is 0.51 days
\citep{her65, pat98}, its position in the orbital period vs.
donor mass plane in Figure \ref{zams_strip_reg100}
indicates $c_1 > 0$.  Thus, we may regard binaries
in the wind phase as ``V~Sge type stars.''

\subsection{Paucity of Progenitor Systems}
     The life time of V~Sge type stars is typically a few to several
times $10^5$ yr, mainly because the time-averaged mass stripping rate
is as high as ${\dot M}_{\rm MS} \sim 10^{-5} M_\sun$~yr$^{-1}$.
If this channel of the WD + MS system produces about four Type Ia
supernovae per millennium in our Galaxy \citep[e.g.,][]{cap99},
we should have a chance to observe at least 
several hundred V~Sge type stars in our Galaxy.
\citet{ste98} listed four V~Sge type stars
in our Galaxy and discussed their similar properties.
Although the masses of the companion stars to the WDs
are not yet clearly identified, their orbital periods fall
in the range of $0.2-0.5$ days, which is very consistent with
the orbital periods predicted by our new scenario 
(see the ``final'' regions in Figs. \ref{zams_evl_con_c3_m110}
-- \ref{zams_final_con_c3_m090}).
However, the total number of V~Sge type stars is too small (by about
two orders of magnitude) to be compatible with the new scenario,
unless 99\% of V~Sge type stars are hidden. 

     The same kind of paucity of the progenitors has been already
pointed out for supersoft X-ray sources (SSXSs) in our Galaxy 
and is attributed to the Galactic interstellar absorption of supersoft
X-rays \citep{dis94}.  Di Stefano \& Rappaport 
also suggested that circumstellar
matter may play some role in the obscuration of X-rays.

\citet{dia95} pointed out that soft X-ray flux of
V~Sge is too weak to be compatible with the typical supersoft X-ray
sources, i.e., at least 2 or 3 orders of magnitude lower than
that of CAL~87, a prototypical SSXS in the LMC.
This obscuration may be explained with the absorption
of X-ray by the stripped matter (or the WD wind itself) and may also 
be related to the observational paucity of the supersoft X-ray sources.

As mentioned in \S 1, \citet{dis03} reported the number of SSXSs in 
four external galaxies from {\it Chandra} data.
They have estimated at least several hundred SSXSs in each galaxy,
many of which are obscured by interstellar absorption.

\subsection{Angular Momentum Loss by Stripped Matter}
The stripped matter is lost from the binary system with
some angular momentum.  In our treatment, we assume
that the specific angular momentum (angular momentum per unit mass)
of the stripped matter is given by equation 
(\ref{stripping_matter_angular_momentum}), that is,
the ablated gas from the companion has the specific angular
momentum there just at the companion's surface.  This assumption
may be too simplified because the stripped matter may get some
angular momentum from the binary motion during its journey.
Here we examine other two cases: one is the same as
the high velocity WD wind, i.e.,
\begin{equation}
\ell_{\rm s} = \left( {{1} \over {1+q}} \right)^2,
\label{fast_stripped_angular_momentum}
\end{equation}
the other is the slow velocity case, i.e.,
\begin{equation}
\ell_{\rm s} = 1,
\label{low_stripped_angular_momentum}
\end{equation}
where the stripped matter gets large angular momentum from 
the binary torque \citep[see][for recent three-dimensional
hydrodynamic calculation]{jah05}.

For the first case of equation (\ref{fast_stripped_angular_momentum}),
we have obtained essentially the same results as in equation
(\ref{stripping_matter_angular_momentum}).
If we adopt the second case of equation 
(\ref{low_stripped_angular_momentum}),
however, we have common envelope formations in a hundred or 
thousand years for $c_1=3$, $M_{2,0}= 5.0~M_\sun$,
and $M_{\rm WD,0}=1.0~M_\sun$ in Figure \ref{zams_evl_con_c3_m100}
regardless of $P_0$.
If we start the evolution with $c_1=3$, $M_{2,0}= 4.0~M_\sun$,
and $M_{\rm WD,0}=1.0~M_\sun$, we obtain SN Ia explosions only for
$P_0=2$--5 days.  These results hardly change even if we increase
the efficiency of mass stripping effect to $c_1=10$. 
This is because too much angular momentum is
removed from the binary for the case of equation
(\ref{low_stripped_angular_momentum})
and it makes the separation shrink drastically
regardless of the $c_1$ value.  Evolutions with $c_1=3$,
$M_{2,0}= 3.5~M_\sun$, and $M_{\rm WD,0}=1.0~M_\sun$ result in
the same final outcome as in equation
(\ref{stripping_matter_angular_momentum}). 

On the other hand, there exist four V~Sge type stars with short orbital
periods of 0.2--0.5 days \citep{ste98}.  Therefore, we conclude 
that the angular momentum loss is much closer to equation
(\ref{stripping_matter_angular_momentum}) 
or (\ref{fast_stripped_angular_momentum})
rather than equation (\ref{low_stripped_angular_momentum}) because
these V~Sge type stars cannot be realized with the large
angular momentum loss like equation (\ref{low_stripped_angular_momentum})
that results in formation of a common envelope.

\subsection{Mass Transfer Rate of Simplified Treatment}
      Our treatment of thermal time scale mass transfer may be
too simplified compared with detailed mass transfer model studied by
\citet{lan00} and \citet{han04}.  Han \& Podsiadlowski compared 
our results based on a simplified model \citep{hknu99} with
their detailed model calculations, and pointed out that the difference
is large for lower mass WDs.  Although we need detailed mass transfer
model to obtain precise SN Ia regions, our treatment has an advantage
of easy and simple estimation for the SN Ia parameter region.
As pointed by \citet{han04}, our SN Ia region thus calculated
may deviate from the realistic one 
for less massive WDs.
However, our SN Ia region is probably not so largely different from
the realistic one for more massive WDs (compare with Fig. 12 of
Hachisu et al. 1999b and Figs. 3 and 5 of Han \& Podsiadlowski 2004).

\section{Concluding Remarks}
Both Cases WIND and CALM originate from the systems with massive
donors, i.e., young population.  It would be important to make
some comparisons with the observational data, such as frequency and
population.  The red hatched regions in Figures
\ref{zams_evl_con_c3_m110}, \ref{zams_evl_con_c3_m100}, and
\ref{zams_evl_con_c3_m090} indicate a region in which the progenitor
explodes at $t_{\rm delay} \le 100$~Myr.  Also the dashed line and the
dotted lines correspond to $t_{\rm delay} =$ 200~Myr and 400~Myr,
respectively.  We see in Figure \ref{birth_rate_population_ms}
that Case WIND and thus SNe Ia/IIn (IIa) are
realized by the very young system with $t_{\rm delay} \lesssim
100-200$~Myr.

If $M_{\rm WD, 0} \lesssim 0.9 ~M_\sun$, we have almost no region of
Case WIND, different from the cases of $M_{\rm WD, 0} \gtrsim
1.0 ~M_\sun$.  If all the WD + MS system with $M_{2,0} 
\gtrsim 3-6 ~M_\sun$ ($c_1=3$), $M_{\rm WD,0} \gtrsim 1.0 ~M_\sun$ 
($M_{1,0} \gtrsim 6.5~M_\sun$), and $P_0 \sim 0.5 -
2$~days produces SNe Ia/IIn (IIa) events
(Table \ref{condition_sn1a_explosion}), the frequency of these events
is estimated to be $\sim 5$\% (including both the WD + MS and
WD + RG systems with their total number ratio of 4:2).

A group of Type IIn SNe such as SNe 1997cy and 1999E show a
very similar spectroscopic and photometric features to SN 2002ic
\citep{wan04, den04, pri07}.
If these are in fact all Type Ia/IIn (IIa) SNe, their
frequency can be estimated to be $\sim 5_{-4}^{+7}$~\% \citep{pri07},
which is consistent with the above estimate.

Type Ia supernovae play a key role in astrophysics, and thus our
progenitor model has important implications. Our model depends
essentially on the parameter of stripping effect, $c_1$,
which depends on the properties of WD winds, such as asphericity,
velocities, and the efficiency of energy conversion.
Also we calculate the mass transfer rate using the simple
approximate binary models.  In order to improve these
parameterization and approximations, we need multi-dimensional
hydrodynamical simulations, which are beyond the scope of
the present study.  In the present approach, we constrain the $c_1$
parameter observationally, and estimate $c_1 \sim 7-8$ and
$c_1 \sim 1.5-10$ from the analysis of V~Sge and RX~J$0513.9-6951$,
respectively.  With keeping in mind the necessity of further
theoretical and observational studies to confirm our new progenitor
systems, we summarize the basic results of our new SN Ia scenario:

(1) Mass-accreting WDs blow an optically thick wind when the mass
transfer rate to the WD exceeds the critical rate of ${\dot M}_{\rm cr}
\sim 1 \times 10^{-6} M_\sun$~yr$^{-1}$.  The WD wind collides with
the secondary's surface and strips off its surface.  If the mass-stripping
effect is efficient enough, the mass transfer rate to the WD is
attenuated and the binary can avoid formation of a common envelope
even for a rather massive secondary. 
Including this mass-stripping effect into our binary evolution model
of the WD + MS systems, we have found a new evolutionary scenario,
in which a companion as massive as 6--$7~M_\sun$ can produce an SN Ia
for a reasonable strength of mass-stripping effect, say $c_1 \sim 3$.

(2) We have followed simplified binary evolutions and obtained the SN Ia
region in the $\log P_0$--$M_{2,0}$ (initial orbital period -- initial
donor mass) plane.   The newly obtained SN Ia region extends to massive
donor masses up to $M_{2,0} \sim 6-7 ~M_\sun$ for $P_0 \sim 0.5-10$ days,
although its extension depends on the strength
of mass-stripping effect, $c_1$, i.e., $M_{2,0} \sim 7-8 ~M_\sun$
for $c_1=10$, $M_{2,0} \sim 5-6 ~M_\sun$ for $c_1=3$,
and $M_{2,0} \sim 4 ~M_\sun$ for $c_1=1$.

(3) We have estimated that the SN Ia birth rate in our Galaxy is
$\nu_{\rm WD+MS} \sim 0.004$~yr$^{-1}$ (for $c_1 = 3$), which is
consistent with the observation.  The rates of young populations,
i.e., $t_{\rm delay} \le 100$~Myr and $t_{\rm delay} \le 200$~Myr,
are about 50\% and 80\% of
the total SN Ia rate of the WD + MS channel.  These short delay times
of SN Ia progenitors are consistent with the recent observational
suggestions that a half of SNe Ia belong to such a very young
population as the delay time of $t_{\rm delay} \sim 10^8$~yr.

(4) Another channel of the WD + RG system shows a broad distribution
of the delay time over 2--3 Gyr \citep{hkn99}, thus the two (WD + MS and
WD + RG) channels yield a bimodality of the delay time distribution.

(5) The stripped-off material is probably distributed on the orbital
plane and forms a massive circumbinary torus (or disk) around SNe Ia.
Such circumstellar matter (CSM) may be consistent with the observed
CSM feature in SN 2006X.
When SN ejecta strongly interact with massive CSM, it can explain
the feature of Type Ia/IIn (IIa) SNe 2002ic and 2006gj.

(6) Three different environments of SN Ia explosions can be specified
by three different states of WDs just at the SN Ia explosion, i.e.,
the optically thick WD wind phase (Case WIND), steady hydrogen
burning phase without optically thick winds from WDs 
(Case CALM), and recurrent nova phase (Case RN).
In Case WIND, SN Ia ejecta strongly interact with massive CSM
like SNe Ia/IIn (IIa) 2002ic and 2005gj because CSM exists near the SN Ia.
The estimated rate of Case WIND is $\sim 5$\% of the total SN Ia rate,
being consistent with the observational estimate.
In Cases CALM and RN, SNe show a normal SN Ia feature because
the CSM is far from the SN but the ejecta may interact with
the CSM in a much later phase.  SN 2006X may be on a border between
Case WIND and Case CALM.

\acknowledgments
We thank Massimo Della Valle and the anonymous referee for their
useful comments.
I.H. and M.K. are grateful to people at the Astronomical Observatory
of Padova and at the Department of Astronomy, University of Padova,
Italy, for their warm hospitality and fruitful discussions,
where we have started and completed this work.  
This research has been supported in part by the Grant-in-Aid for
Scientific Research (16540211, 18104003, 18540231)
of the Japan Society for the Promotion of Science,
and by the NSF under grant PHY99-07949.
We would like to thank stimulated discussion
at the Santa Barbara workshop ``Paths to Exploding Stars:
Accretion and Explosion'' (19-23 March 2007).

\end{document}